\documentclass[12pt]{article}

\usepackage[totalwidth=460truept,totalheight=610truept]{geometry}
\usepackage{amsfonts,latexsym,amssymb,amsmath,graphicx,accents,slashed,subfigure}
\usepackage{latexsym}
\usepackage[hidelinks]{hyperref}

\global\arraycolsep=1truept

\begin{document}

\null

\vskip1truecm

\begin{center}
{\huge \textbf{The Quest For}}

\vskip.65truecm

{\huge \textbf{Purely Virtual Quanta:}}

\vskip.65truecm

{\huge \textbf{Fakeons Versus}}

\vskip.8truecm

{\huge \textbf{Feynman-Wheeler Particles}}

\vskip1truecm

\textsl{Damiano Anselmi}

\vskip .1truecm

\textit{Dipartimento di Fisica \textquotedblleft Enrico
Fermi\textquotedblright , Universit\`{a} di Pisa}

\textit{Largo B. Pontecorvo 3, 56127 Pisa, Italy}

\textit{and INFN, Sezione di Pisa,}

\textit{Largo B. Pontecorvo 3, 56127 Pisa, Italy}

damiano.anselmi@unipi.it

\vskip2truecm

\textbf{Abstract}
\end{center}

The search for purely virtual quanta has attracted interest in the past. We
consider various proposals and compare them to the concept of fake particle,
or \textquotedblleft fakeon\textquotedblright . In particular, the
Feynman-Wheeler propagator, which amounts to using the Cauchy principal
value inside Feynman diagrams, violates renormalizability, unitarity and
stability, due to the coexistence of the prescriptions $\pm i\epsilon $. We
contrast the Feynman, fakeon and Feynman-Wheeler prescriptions in ordinary
as well as cut diagrams. The fakeon does not have the problems of the
Feynman-Wheeler propagator and emerges as the correct concept of purely
virtual quantum. It allows us to make sense of quantum gravity at the
fundamental level, and places it on an equal footing with the standard
model. The resulting theory of quantum gravity is perturbative up to an
incredibly high energy.

\vfill\eject

\section{Introduction}

\label{intro}\setcounter{equation}{0}

A normal particle can be real or virtual, depending on whether it is on
shell or off shell. For a variety of reasons, it is interesting to consider
the possibility of having quanta that are purely virtual. This means quanta
that propagate inside the scattering processes, but cannot be directly
observed, because they do not belong to the physical spectrum, i.e. the set
of asymptotic states.

Sings of interest for purely virtual particles have been present in the
literature for a long time, at the classical and quantum levels. Dirac
considered the Abraham-Lorentz force in classical electrodynamics, which
effectively describes the recoil on an accelerated pointlike electric charge
due to the emission of radiation. It is well-known that, under certain
assumptions, this effect can be described by means of a higher-derivative
equation, which has undesirable runaway solutions. Dirac \textquotedblleft
virtualized\textquotedblright\ the runaway solutions by trading them for
violations of microcausality \cite{dirac,jackson}. A similar trick can be
applied to higher-derivative gravity \cite{zia}, still at the classical
level. Feynman and Wheeler studied a version of classical electrodynamics
where the Green function is half the sum of the retarded and advanced
potentials \cite{FW} and recovered causality by means of an involved
emitter-absorber theory. The same Green function, that is to say, the Cauchy
principal value of the unprescribed propagator $1/(p^{2}-m^{2})$, appears in
many contexts. Differently from the retarded and advanced potentials and the
Feynman propagator, it does not contain the on shell $\delta $ function, so
it may be viewed as a candidate to describe purely virtual particles. As
such, Bollini and Rocca in ref. \cite{bollini} and later Plastino and Rocca
in ref. \cite{rocca} studied it at the quantum level and claimed they could
make some sense out of it.

The interest for purely virtual quanta was not misplaced, but, for a variety
of reasons that we explain in this paper, the principal value of $%
1/(p^{2}-m^{2})$ cannot be the right answer at the quantum level, because it
generates serious problems when it is used inside Feynman diagrams. First,
it violates the locality of counterterms by generating nonlocal ultraviolet
divergences similar to those of ref. \cite{ugo}, whose removal destroys the
basic structure of the theory. Second, it generates imaginary parts that are
inconsistent with the optical theorem and unitarity. Third, it violates
stability, since the ones that are normally known as pseudothresholds become
true thresholds, which leads to processes where the incoming energy is equal
to a difference of frequencies, rather than the sum.

The problems are due to the coexistence of the $\pm i\epsilon $
prescriptions within the same diagram. In ref. \cite{ugo} Aglietti and the
current author showed that for finite $\epsilon $, these types of diagrams
have nonlocal divergent parts in Minkowski spacetime. Although the analysis
of \cite{ugo} was performed in higher derivative theories, its main
conclusions extend to infinitesimal widths and the principal-value
propagator.

To fix a bit of terminology, the principal-value propagator will be called
Feynman-Wheeler (FW)\ propagator, when it is used inside Feynman diagrams.
The degrees of freedom it propagates will be called FW\ particles.

In the end, the FW\ particles lack a consistent physical interpretation and
are even problematic at the mathematical level, due to the nonlocal
divergences. The correct notion of purely virtual quantum turns out to be
the fakeon \cite{LWgrav,fakeons}, which is encoded in a new prescription to
quantize the poles of a free propagator. Although the fakeon propagator
tends to the principal value of $1/(p^{2}-m^{2})$ in the classical
free-field limit, it radically differs from it at the quantum level and when
self-interactions are turned on \cite{FLRW}. Not surprisingly, the nature of
a purely virtual quantum is... purely quantum, so the classical limit is not
enough to infer the quantum nature of the fakeon.

The fakeons do not violate stability, because the pseudothresholds play no
significant role. Actually, the thresholds associated with a Feynman diagram
that involves fakeons are exactly the ones that are found by means of the
usual Feynman prescription. The locality of counterterms is still valid,
because the divergent part of a diagram coincides with the one of its
Euclidean version \cite{fakeons}. Finally, unitarity is fulfilled, since
the\ imaginary parts of the amplitudes do not receive contributions from the
thresholds of the processes that involve fakeons. This means that such
processes have zero chances to turn the fakeons into real particles, which
is the reason why the fakeons remain virtual and do not belong to the set of
asymptotic states.

It is worth to stress that the properties just listed hold at the
fundamental level, not just at the effective one. In no other case we can
really drop a particle from the physical spectrum (letting aside gauge
artifacts, such as the Faddeev-Popov ghosts and the unphysical modes of the
gauge fields). In particular, we cannot drop unstable particles. For
example, although the muon decays, it can be observed before it does. In the
end, only the fakeon fulfills the requirements to be called a purely virtual
quantum, which is possible because it is a thoroughly new concept.

An important application of the fakeon prescription is that it allows us to
make sense of quantum gravity as a quantum field theory \cite{LWgrav}, by
reconciling unitarity and renormalizability. The resulting theory propagates
the graviton, a massive scalar field $\phi $, which can be either a physical
particle or a fakeon, and a massive spin-2 fakeon $\chi _{\mu \nu }$. The
masses $m_{\phi }$ and $m_{\chi }$ are free parameters. Presumably, their
values are smaller than the Planck mass, or even much smaller. If so, the
quantum gravity theory obtained from the fakeon quantization turns out to be
perturbative up to an unbelievably high energy, an energy so high that we
think it deserves to be named \textquotedblleft God's
energy\textquotedblright . At the same time, it leads to the troubling
scenario of an infinite desert with no new physics between the Planck scale
and God's energy.

Something that the Dirac virtualization, the FW idea and the fakeon have in
common is that they all lead to violations of causality at very small
distances (in different ways in the three cases) \cite{causality,FLRW}.
Physically, the violation of microcausality is not a high price to pay,
since we do not have arguments in favor of absolute causality. It must also
be recalled that it is not straightforward to define the concept of
causality in quantum field theory, even in flat space\footnote{%
See section 6.1 of \cite{diagrammar} for an illuminating discussion.}, since
it is hard to accurately locate spacetime points when we describe on-shell
particles by means of relativistic wave packets. Both the Bogoliubov
condition \cite{bogoliubov} and the Lehmann-Symanzik-Zimmermann requirement
that fields commute at spacelike separated points \cite{LSZ} are off shell
and do not easily translate into properties of the $S$ matrix.

Clearly, the ideas of causality and time ordering make sense only as long as
the notion of time makes sense. At the experimental level, the shortest time
interval we can measure directly is 10$^{-17}$-10$^{-18}$ seconds \cite%
{laserpulses}. The theory of quantum gravity based on fakeons predicts that
time loses meaning at distances smaller than the Compton wavelength $%
1/m_{\chi }$ of the fakeon $\chi _{\mu \nu }$ (or the larger between $%
1/m_{\chi }$ and $1/m_{\phi }$, if both $\chi _{\mu \nu }$ and $\phi $ are
fakeons). This means 10$^{-36}$s, i.e. almost twenty orders of magnitude
away from our present accuracy, if the masses $m_{\chi }$, $m_{\phi }$ are
assumed to be around 10$^{12}$GeV. The simplest attempts we can think of to
amplify the effect do not lead too far. Actually, in some cases the universe
conspires to recover causality for free \cite{marino}. In the classical
limit, we may have to downgrade the violation of causality to an unusual
form of the equations of motion and a fuzziness of their solutions \cite%
{FLRW,marino}, which still lead to predictions that can be tested
experimentally. Finally, we stress that even if time loses sense at small
scales, scattering processes in momentum space make sense to arbitrarily
high energies.

The paper is organized as follows. In section \ref{nonlocal} we calculate
the nonlocal divergent parts of the bubble diagrams involving
Feynman-Wheeler particles. In section \ref{finite} we evaluate their finite
parts. In section \ref{unitarity} we analyze their imaginary parts and show
that they violate the optical theorem. In section \ref{stability} we discuss
the problems of the FW particles with stability. In section \ref{fakeons} we
compute the bubble diagrams with circulating fakeons and show that the
fakeon prescription is consistent with the locality of counterterms,
unitarity and stability. In section \ref{three} we extend the analysis to
three and two spacetime dimensions, which confirm the properties found in
four dimensions. In section \ref{remarks} we discuss the impact of the
fakeon idea on the perturbative nature of quantum gravity. Section \ref%
{conclusions} contains the conclusions.

\section{Bubble diagram with FW\ particles: nonlocal divergent part}

\setcounter{equation}{0}\label{nonlocal}

In this section we show that the FW propagator%
\begin{equation}
\mathcal{P}\frac{1}{p^{2}-m^{2}},  \label{FWprop}
\end{equation}%
where $\mathcal{P}$ denotes the Cauchy principal value, leads to nonlocal
ultraviolet divergences when it is used inside Feynman diagrams. Note that
in this paper we do not attach factors of $i$ to the vertices and the
propagators.

Specifically, we consider the bubble diagram where one virtual particle is
quantized by means of the Feynman $+i\epsilon $ prescription and the other
virtual particle has the propagator (\ref{FWprop}). The loop integral reads%
\begin{equation*}
\Sigma _{\text{FW}}(p)=\int\limits_{k_{s}\leqslant \Lambda }\frac{\mathrm{d}%
^{3}\mathbf{k}}{(2\pi )^{3}}\int\limits_{-\infty }^{+\infty }\frac{\mathrm{d}%
k_{0}}{2\pi }\hspace{0.01in}\frac{1}{(p-k)^{2}-m_{1}^{2}+i\epsilon }\,%
\mathcal{P}\frac{1}{k^{2}-m_{2}^{2}},
\end{equation*}%
where $k=(k^{0},\mathbf{k})$, $k_{s}=|\mathbf{k}|$ and $\Lambda $ is a
cutoff. We take different masses $m_{1}\neq m_{2}$, because the nonlocal
divergent part that we want to compute vanishes for $m_{1}=m_{2}$.

We anticipate that, apart from the locality of counterterms, which is lost,
other common properties continue to hold. For example, although the cutoff $%
\Lambda $ breaks Lorentz invariance, the breaking does not affect the finite
parts and the logarithmic divergences we find below. In section \ref{finite}
we switch to the dimensional regularization, which is manifestly Lorentz
invariant.

Writing%
\begin{equation*}
\mathcal{P}\frac{1}{k^{2}-m_{2}^{2}}=\frac{1}{2}\left( \frac{1}{%
k^{2}-m_{2}^{2}-i\epsilon }+\frac{1}{k^{2}-m_{2}^{2}+i\epsilon }\right) ,
\end{equation*}%
we can decompose $\Sigma _{\text{FW}}(p)$ as the sum%
\begin{equation}
\Sigma _{\text{FW}}(p)=\frac{1}{2}\Big(\Sigma (p)+\Sigma ^{\prime }(p)\Big),
\label{sfw}
\end{equation}%
where $\Sigma (p)$ is the usual self-energy, with two propagators quantized 
\`{a} la Feynman [its expression being given by formula (\ref{sigmap}) of
section \ref{fakeons}], and 
\begin{equation*}
\Sigma ^{\prime }(p)=\int\limits_{k_{s}\leqslant \Lambda }\frac{\mathrm{d}%
^{3}\mathbf{k}}{(2\pi )^{3}}\int\limits_{-\infty }^{+\infty }\frac{\mathrm{d}%
k_{0}}{2\pi }\hspace{0.01in}\frac{1}{(p-k)^{2}-m_{1}^{2}+i\epsilon }\,\frac{1%
}{k^{2}-m_{2}^{2}-i\epsilon }.
\end{equation*}%
Since $\Sigma (p)$ does not contain nonlocal divergent parts, we omit it for
the moment and focus on $\Sigma ^{\prime }(p)$. The crucial point is that $%
\Sigma ^{\prime }(p)$ involves both prescriptions $\pm i\epsilon $. The
nonlocal divergent part is due to the conflict between them.

Writing $p=(p^{0},\mathbf{p})$, we apply the residue theorem to evaluate the
integral on the loop energy $k_{0}$, which gives%
\begin{equation}
\Sigma ^{\prime }(p)=\frac{i}{4}\int\limits_{k_{s}\leqslant \Lambda }\frac{%
\mathrm{d}^{3}\mathbf{k}}{(2\pi )^{3}}\frac{1}{\omega _{1}\omega _{2}}\left( 
\hspace{0.01in}\frac{1}{p^{0}-\omega _{1}+\omega _{2}+i\epsilon }-\frac{1}{%
p^{0}+\omega _{1}-\omega _{2}-i\epsilon }\right) \,,  \label{sp}
\end{equation}%
where $\omega _{1}=\sqrt{k_{s}^{2}+p_{s}^{2}-2k_{s}p_{s}u+m_{1}^{2}}$, $%
\omega _{2}=\sqrt{k_{s}^{2}+m_{2}^{2}}$, $p_{s}=|\mathbf{p}|$ and $u=\cos
\theta $, $\theta $ being the angle between $\mathbf{k}$\ and $\mathbf{p}$.

Let us start from the case $p^{2}\geqslant 0$ and choose a reference frame
where $p_{s}=0$. Then, it is easy to check that%
\begin{equation}
(\omega _{1}-\omega _{2})^{2}\leqslant (m_{1}-m_{2})^{2}.  \label{omom}
\end{equation}%
If we assume $p^{2}>(m_{1}-m_{2})^{2}$, (\ref{omom}) gives $%
(p^{0})^{2}>(\omega _{1}-\omega _{2})^{2}$, which means that the $\pm
i\epsilon $ prescriptions in (\ref{sp}) are unnecessary. Then we obtain%
\begin{equation}
\Sigma ^{\prime }(p)=\frac{i}{8\pi ^{2}}\int\limits_{0}^{\Lambda }\frac{%
k_{s}^{2}\mathrm{d}k_{s}}{\omega _{1}\omega _{2}}\left( \frac{1}{%
p^{0}-\omega _{1}+\omega _{2}}-\hspace{0.01in}\frac{1}{p^{0}+\omega
_{1}-\omega _{2}}\right) .  \label{spr}
\end{equation}%
It is easy to check that the integrand behaves as%
\begin{equation*}
\frac{m_{1}^{2}-m_{2}^{2}}{(p^{0})^{2}k_{s}}+\mathcal{O}\left( \frac{1}{%
k_{s}^{3}}\right)
\end{equation*}%
for large $k_{s}$, which means that $\Sigma ^{\prime }(p)$ has the nonlocal
ultraviolet divergence%
\begin{equation}
\Sigma _{\text{div}}^{\prime }(p)=\frac{i\ln \Lambda ^{2}}{(4\pi )^{2}}\frac{%
m_{1}^{2}-m_{2}^{2}}{p^{2}}.  \label{diva}
\end{equation}

To check Lorentz invariance, we repeat the calculation for $p^{2}>0$ without
taking $p_{s}=0$. In that case, the inequality (\ref{omom}) non longer
holds, in general. However, we have $(p^{0})^{2}-(\omega _{1}-\omega
_{2})^{2}=p^{2}+p_{s}^{2}(1-u^{2})+\mathcal{O}\left( 1/k_{s}\right) $, so
when $k_{s}$ is sufficiently large we do have $(p^{0})^{2}>(\omega
_{1}-\omega _{2})^{2}$ and we can repeat the arguments above, ignoring the $%
\pm i\epsilon $ prescriptions. At the end, we find the same nonlocal
divergent part (\ref{diva}).

Finally, if $p^{2}<0$ we take $p^{0}=0$. We need to compute 
\begin{equation*}
\Sigma ^{\prime }(p)=\frac{i}{8\pi ^{2}}\int\limits_{0}^{\Lambda }\frac{%
k_{s}^{2}\mathrm{d}k_{s}}{\omega _{1}\omega _{2}}\int_{-1}^{1}\hspace{0.01in}%
\frac{\mathrm{d}u}{\omega _{2}-\omega _{1}+i\epsilon }.
\end{equation*}%
For large $k_{s}$, the divergent part is given by%
\begin{eqnarray}
\Sigma _{\text{div}}^{\prime }(p) &=&\frac{i}{8\pi ^{2}}\int\limits_{0}^{%
\Lambda }\mathrm{d}k_{s}\int_{-1}^{1}\mathrm{d}u\left[ \frac{1}{%
p_{s}(u+i\epsilon )}+\frac{m_{1}^{2}-m_{2}^{2}+p_{s}^{2}(1+u^{2})}{%
2k_{s}p_{s}^{2}(u+i\epsilon )^{2}}\right]  \notag \\
&=&\frac{\Lambda }{8\pi p_{s}}-\frac{i(m_{1}^{2}-m_{2}^{2})}{(4\pi
)^{2}p_{s}^{2}}\ln \Lambda ^{2}  \label{linear}
\end{eqnarray}%
(after suitably rescaling $\epsilon $). The logarithmic divergence agrees
with (\ref{diva}). In addition, we have a nonlocal linear divergence. We do
not attach a particular meaning to it, because it depends on the regulator
and indeed disappears using the dimensional regularization (see next
section).

The nonlocal divergent parts we have just calculated are very similar to
those found in ref. \cite{ugo} in higher-derivative theories with finite
widths $\epsilon $. They have the same origin: the coexistence of
propagators with both positive and negative widths within the same Feynman
diagram.

Note that the nonlocal divergent part (\ref{diva}) cancels out in the bubble
diagram%
\begin{equation}
\Sigma _{\text{FW-FW}}(p)=\mathcal{P}\int \frac{\mathrm{d}^{D}k}{(2\pi )^{D}}%
\hspace{0.01in}\frac{1}{(p-k)^{2}-m_{1}^{2}}\,\frac{1}{k^{2}-m_{2}^{2}}=%
\frac{1}{2}\Big(\Sigma _{\text{FW}}(p)+\Sigma _{\text{FW}}^{\ast }(p)\Big)
\label{sfwfw}
\end{equation}%
made of two FW propagators, since the result must be symmetric under the
exchange of $m_{1}$ and $m_{2}$. For future purposes, we have written (\ref%
{sfwfw}) in arbitrary dimension $D$.

Since (\ref{diva}) does not cancel out in mixed bubble diagrams $\Sigma _{%
\text{FW}}(p)$, it must be subtracted away by means of a nonlocal
counterterm, which destroys the locality of the theory. In the end, we must
conclude that a theory propagating FW particles is mathematically
unacceptable.

\section{Bubble diagram with FW\ particles: finite part}

\setcounter{equation}{0}\label{finite}

In this section we work out the finite part of $\Sigma ^{\prime }(p)$, which
allows us to highlight the physical problems of the theories that propagate
FW\ particles.

We start by taking $m_{2}=0$. For $p^{2}>0$ we choose a reference frame
where $p_{s}=0$. Using (\ref{sp}) and denoting $m_{1}$ by $m$, we obtain%
\begin{equation}
i(4\pi )^{2}\Sigma ^{\prime }(p)=-\frac{m^{2}}{p^{2}}\ln \frac{4\Lambda ^{2}%
}{m^{2}}-\frac{p^{2}-m^{2}}{p^{2}}\ln \frac{p^{2}-m^{2}+i\epsilon }{p^{2}}.
\label{finitesopra}
\end{equation}%
Note the negative imaginary part 
\begin{equation}
2\mathrm{Im}[-i\Sigma ^{\prime }(p)]=-\frac{m^{2}-p^{2}}{8\pi p^{2}}\theta
(m^{2}-p^{2}),  \label{im1}
\end{equation}%
which is problematic for unitarity (see next section).

To switch to the case of generic masses, it is convenient to use the
dimensional regularization \cite{dimreg}. The calculation by means of
Feynman parameters exhibits some unexpected features. Indeed, we cannot use
the Feynman parameters for $\Sigma ^{\prime }(p)$ as we would ordinarily do.
We must first change the sign of one propagator to have $+i\epsilon $ in
both of them. After a translation of the loop momentum $k$, we get 
\begin{eqnarray*}
\Sigma ^{\prime }(p) &=&-\int \frac{\mathrm{d}^{D}k}{(2\pi )^{D}}\hspace{%
0.01in}\frac{1}{(p-k)^{2}-m_{1}^{2}+i\epsilon }\,\frac{1}{%
-k^{2}+m_{2}^{2}+i\epsilon } \\
&=&-\int_{0}^{1}\mathrm{d}x\int \frac{\mathrm{d}^{D}k}{(2\pi )^{D}}\frac{1}{%
\left[ i\epsilon -(1-2x)k^{2}+p^{2}\frac{x(1-x)}{1-2x}%
-m_{1}^{2}x+m_{2}^{2}(1-x)\right] ^{2}},
\end{eqnarray*}%
where $D$ denotes the continued spacetime dimension. At this point, we break
the $x$ integral into the sum of two pieces, the integral on $0\leqslant
x\leqslant 1/2$ and the integral on $1/2\leqslant x\leqslant 1$. The two are
defined by opposite $i\epsilon $ prescriptions, since the coefficients of $%
k^{2}$ have opposite signs.

We can check the method in the simple case $m_{2}=0$, $m_{1}=m$, where we
get, after expanding around $D=4$,%
\begin{eqnarray*}
i(4\pi )^{2}\Sigma ^{\prime }(p) &=&-\left[ \frac{2}{4-D}+2-\gamma _{E}+\ln
(4\pi )\right] \frac{m^{2}p^{2}}{(p^{2})^{2}+\epsilon ^{2}}+\frac{1}{2}\ln
\left( (p^{2})^{2}+\epsilon ^{2}\right) \\
&&+\frac{m^{2}\ln m^{2}}{p^{2}-i\epsilon }-\frac{m^{2}}{2}\left[ \frac{\ln
(p^{2}+i\epsilon )}{p^{2}+i\epsilon }+\frac{\ln (p^{2}-i\epsilon )}{%
p^{2}-i\epsilon }\right] \\
&&-\frac{p^{2}-m^{2}+i\epsilon }{p^{2}+i\epsilon }\ln (p^{2}-m^{2}+i\epsilon
).
\end{eqnarray*}%
This result agrees with (\ref{finitesopra}) for $p^{2}>0$ upon the cutoff
identification%
\begin{equation}
\ln \Lambda ^{2}=\frac{2}{4-D}+2-\gamma _{E}+\ln \pi .  \label{cuto}
\end{equation}

With generic masses, we proceed as follows. After breaking the $x$ integral
into the sum of the integrals on $0\leqslant x\leqslant 1/2$ and $%
1/2\leqslant x\leqslant 1$, we rescale $\epsilon $ and convert the second
integral into another integral on $0\leqslant x\leqslant 1/2$ by means of
the change of variables $x\rightarrow 1-x$. So doing, we get%
\begin{equation}
\Sigma ^{\prime }(p)=U(p^{2}+i\epsilon
,m_{1}^{2},m_{2}^{2})-U(p^{2}-i\epsilon ,m_{2}^{2},m_{1}^{2}),
\label{spfinite}
\end{equation}%
where%
\begin{equation}
U(a,b,c)=\frac{i\Gamma \left( \frac{4-D}{2}\right) }{(4\pi )^{D/2}}%
\int_{0}^{1/2}\mathrm{d}x\hspace{0.01in}(1-2x)^{2-D}\left[
ax(1-x)-(1-2x)(bx-c(1-x))\right] ^{(D-4)/2}.  \label{uabc}
\end{equation}%
The $x$ integration is relatively straightforward and, after the expansion
around $D=4$, we obtain%
\begin{equation}
i(4\pi )^{2}U(p^{2},m_{1}^{2},m_{2}^{2})=\frac{v_{+}}{2p^{2}}\left( \ln 
\frac{4\Lambda ^{2}}{m_{2}^{2}}-z\ln \frac{1+z}{1-z}\right) ,  \label{u}
\end{equation}%
where $\ln \Lambda ^{2}$ is defined as in (\ref{cuto}) and%
\begin{equation}
z=\frac{\sqrt{u_{+}u_{-}}}{v_{+}},\qquad u_{\pm }=(m_{1}\pm
m_{2})^{2}-p^{2},\qquad v_{\pm }=p^{2}\mp m_{1}^{2}\pm m_{2}^{2}.
\label{upm}
\end{equation}

Since $\Sigma ^{\prime }$ turns into its complex conjugate under the
replacement $m_{1}\leftrightarrow m_{2}$, we can assume, with no loss of
generality, that $m_{1}>m_{2}$. If we define 
\begin{equation*}
x=\frac{\sqrt{|u_{+}u_{-}|}}{v_{+}},\qquad y=\frac{\sqrt{|u_{+}u_{-}|}}{v_{-}%
},\qquad z^{\prime }=\frac{\sqrt{u_{+}u_{-}}}{v_{-}},
\end{equation*}%
we find the table 
\begin{equation}
\begin{tabular}{c|c|c|c}
$p^{2}$ range & $\left. z\ln \frac{1+z}{1-z}\right\vert _{p^{2}\rightarrow
p^{2}+i\epsilon }$ & $\left. z^{\prime }\ln \frac{1+z^{\prime }}{1-z^{\prime
}}\right\vert _{p^{2}\rightarrow p^{2}-i\epsilon }$ &  \\[0.5em] \hline
&  &  &  \\[-1em] 
$u_{+}<0$ & $x\ln \left( \frac{1+x}{1-x}\right) $ & $y\ln \left( \frac{1+y}{%
1-y}\right) $ & $0<x<1,$ $0<y<1$ \\[0.5em] 
$-v_{+}<0<u_{+}$ & $-2x\arctan (x)$ & $-2y\arctan (y)$ & $x>0,$ $y>0$ \\%
[0.5em] 
$u_{-}<0<-v_{+}$ & $-2x\arctan (x)-2\pi x$ & $-2y\arctan (y)$ & $x<0,$ $y>0$
\\[0.5em] 
$-p^{2}<0<u_{-}$ & $x\ln \left( \frac{1+x}{1-x}\right) -2i\pi x$ & $y\ln
\left( \frac{1+y}{1-y}\right) $ & $-1<x<0,$ $0<y<1$ \\[0.5em] 
$-v_{-}<0<-p^{2}$ & $x\ln \left( \frac{1+x}{x-1}\right) -i\pi x$ & $y\ln
\left( \frac{1+y}{y-1}\right) +i\pi y$ & $x<-1,$ $y>1$ \\[0.5em] 
$0<-v_{-}$ & $x\ln \left( \frac{1+x}{x-1}\right) -i\pi x$ & $y\ln \left( 
\frac{1+y}{y-1}\right) +i\pi y$ & $x<-1,$ $y<-1$%
\end{tabular}
\label{sempla}
\end{equation}

Finally, if we use (\ref{sigmap}) for $\Sigma (p)$, formula (\ref{sfw}) gives%
\begin{equation}
\Sigma _{\text{FW}}(p)=\frac{1}{2}\Big(V(p^{2}+i\epsilon
,m_{1}^{2},m_{2}^{2})+U(p^{2}+i\epsilon
,m_{1}^{2},m_{2}^{2})-U(p^{2}-i\epsilon ,m_{2}^{2},m_{1}^{2})\Big),
\label{sfwres}
\end{equation}%
where $V(a,b,c)$ is given by formula (\ref{vabc}) of section \ref{fakeons}.
In the case of $\Sigma _{\text{FW-FW}}$, formula (\ref{sfwfw}) gives 
\begin{eqnarray}
\Sigma _{\text{FW-FW}}(p) &=&\frac{1}{4}\Big(V(p^{2}+i\epsilon
,m_{1}^{2},m_{2}^{2})+U(p^{2}+i\epsilon
,m_{1}^{2},m_{2}^{2})-U(p^{2}-i\epsilon ,m_{2}^{2},m_{1}^{2})  \label{sfwfwf}
\\
&&-V(p^{2}-i\epsilon ,m_{1}^{2},m_{2}^{2})-U(p^{2}-i\epsilon
,m_{1}^{2},m_{2}^{2})+U(p^{2}+i\epsilon ,m_{2}^{2},m_{1}^{2})\Big).  \notag
\end{eqnarray}

Now we analyze the meaning of these results.

\section{Unitarity}

\setcounter{equation}{0}\label{unitarity}

In this section, we study the imaginary parts of the amplitudes and show
that the theory propagating FW\ particles is not consistent with the optical
theorem.

The optical theorem is another way to express the unitarity of the $S$
matrix. Writing $S=1+iT$, the identity $S^{\dag }S=1$ becomes $-iT+iT^{\dag
}=T^{\dag }T$ and can be rephrased diagrammatically by means of cut
diagrams, which are diagrams divided into two parts by a cut that crosses
internal legs \cite{cut,diagrammar}. One side of the cut is due to the
factor $T$ of $T^{\dag }T$. There, the diagram is unshadowed and the
vertices and propagators are those given by the usual Feynman rules (once
the usual factors of $i$ are restored in vertices and propagators). The
other side of the cut is due to the factor $T^{\dag }$ of $T^{\dag }T$. It
is shadowed and the vertices and propagators are the complex conjugates of
those given by the Feynman rules. Finally, \textquotedblleft cut
propagators\textquotedblright\ account for the legs crossed by the cut. The
cut lines represent the product of $T^{\dag }$ and $T$ in $T^{\dag }T$.

The cut diagrams encode the imaginary part of the amplitude and the cross
section for the production of the particles circulating in the loop. This
means that, beyond appropriate thresholds, the virtual particles of the loop
may turn into real particles.

\begin{figure}[t]
\begin{center}
\includegraphics[width=12truecm]{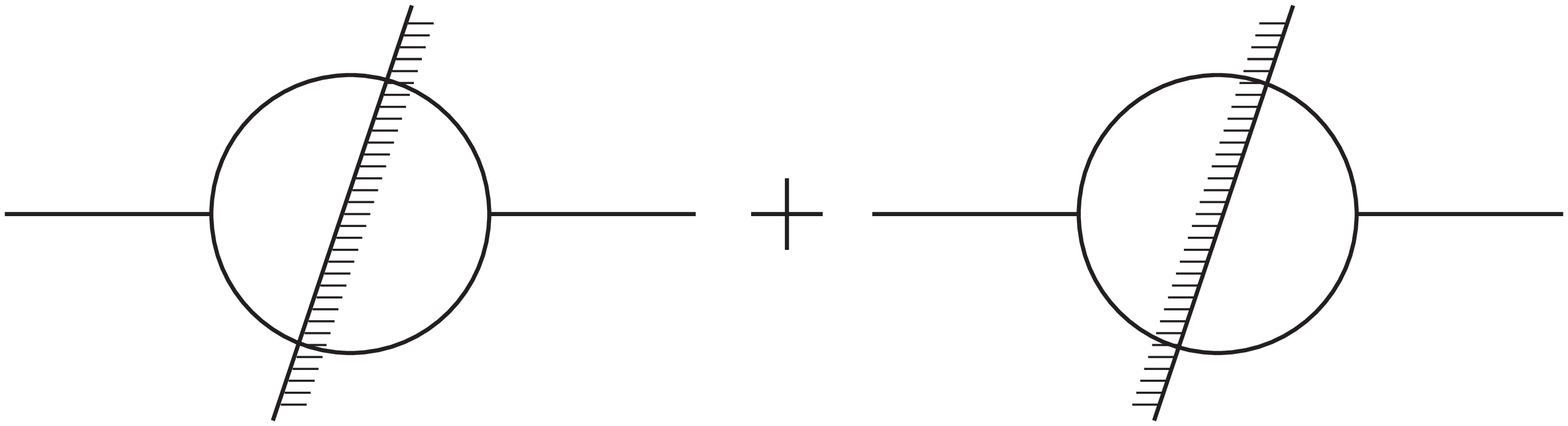}
\end{center}
\caption{Cut diagrams that give the imaginary part of $-2i$ times the bubble
diagram}
\label{cut}
\end{figure}

\begin{figure}[t]
\begin{center}
\includegraphics[width=13truecm]{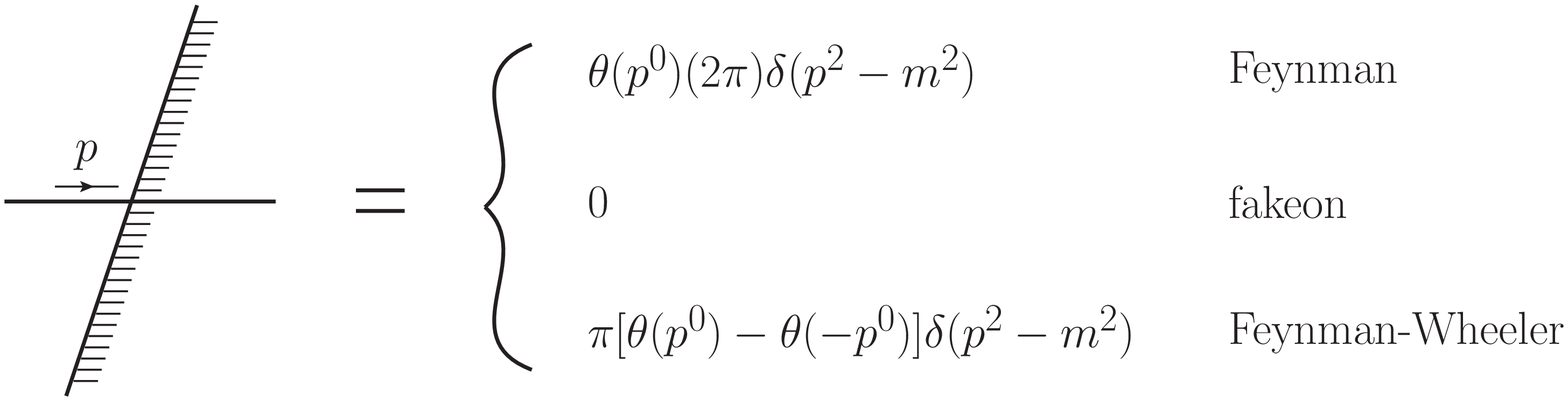}
\end{center}
\caption{Cut propagators in the various cases}
\label{cutprop}
\end{figure}

Let us start from the usual bubble diagram $\Sigma (p)$, where both
propagators are defined by means of the Feynman prescription. There, the
true threshold is $p^{2}=(m_{1}+m_{2})^{2}$ (i.e. $u_{+}=0$) and above it,
the imaginary part $\mathrm{Im}(-i\Sigma )$ is nontrivial. From formula (\ref%
{hollik}) of section \ref{fakeons} we find%
\begin{equation}
2\mathrm{Im}(-i\Sigma )=\frac{\sqrt{u_{+}u_{-}}}{8\pi p^{2}}\theta (-u_{+}).
\label{imma}
\end{equation}%
This result obeys the optical theorem, which states in this case that 2$%
\mathrm{Im}(-i\Sigma )$ is equal to the sum of the cut diagrams shown in
fig. \ref{cut}, where the cut propagator is the top one of fig. \ref{cutprop}%
. It means that for $p^{2}>(m_{1}+m_{2})^{2}$ the incoming particle can
decay into the virtual particles circulating in the loop, which are then
turned into real particles.

For future use, we recall that one cut diagram is obtained by replacing the
two internal propagators of $\Sigma (p)$ by the cut propagators%
\begin{equation}
(2\pi )\theta (p^{0}-k^{0})\delta ((p-k)^{2}-m_{1}^{2}),\qquad (2\pi )\theta
(k^{0})\delta (k^{2}-m_{2}^{2}).  \label{tt}
\end{equation}%
The other cut diagram is obtained by flipping the signs of the arguments of
the $\theta $ functions.

The Landau equations \cite{landaueq} provide a systematic method to identify
the potential thresholds of a loop integral, where the amplitude may be
nonanalytic. It is well known that in the case of $\Sigma (p)$ the Landau
equations give a second potential threshold, which is $%
p^{2}=(m_{1}-m_{2})^{2}$ (i.e. $u_{-}=0$). However, that threshold is not
associated with any pinching singularities of the integral $\Sigma (p)$,
which means that it is not a true threshold. For this reason, it is commonly
called pseudothreshold.

When we consider $\Sigma ^{\prime }(p)$, we find that the two potential
thresholds exchange roles and $p^{2}=(m_{1}-m_{2})^{2}$ becomes the true
threshold, while $p^{2}=(m_{1}+m_{2})^{2}$ becomes a pseudothreshold.
Moreover, the imaginary part of $-i\Sigma ^{\prime }$ is nonvanishing below
its true threshold, which means for $p^{2}<(m_{1}-m_{2})^{2}$, not above it.
Specifically, using (\ref{sempla}), formulas (\ref{spfinite}) and (\ref{u})
allow us to derive%
\begin{equation}
2\mathrm{Im}(-i\Sigma ^{\prime })=-\frac{\sqrt{u_{+}u_{-}}}{8\pi p^{2}}%
\theta (u_{-}).  \label{ims}
\end{equation}

The troubles with FW particles are evident from this formula. Indeed, taking 
$\Sigma ^{\prime }=iT$, the optical theorem $-iT+iT^{\dag }=T^{\dag }T$
interprets the left-hand side of (\ref{ims}) as a forward scattering
amplitude and implies that the right-hand side is the total cross section
for the production of the virtual-turned-into-real particles in all final
states. In particular, since $T^{\dag }T\geqslant 0$, the right-hand side of
(\ref{ims}) should be nonnegative, which is evidently false for $p^{2}>0$.
Basically, (\ref{ims}) says that at $p^{2}>0$ an incoming particle has a
negative probability to decay into the final particles! Moreover, (\ref{ims}%
) is singular for $p^{2}=0$, which means that for $p^{2}\rightarrow 0^{\pm }$
the cross section is infinite. Such a singularity is clearly absent in the
usual case, as formula (\ref{imma}) shows.

The result (\ref{ims}) also admits an interpretation in terms of cut
diagrams. Precisely, for $p^{2}>0$ the right-hand side of (\ref{ims}) is
still equal to the sum of the cut diagrams of fig. \ref{cut}, but the left
diagram has cut propagators%
\begin{equation}
(2\pi )\theta (p^{0}-k^{0})\delta ((p-k)^{2}-m_{1}^{2}),\qquad -(2\pi
)\theta (-k^{0})\delta (k^{2}-m_{2}^{2}),  \label{tt2}
\end{equation}%
while the right diagram is obtained from the left one by flipping the signs
of the arguments of the $\theta $ functions.

The important point is that the sign of the argument of second $\theta $
function of formula (\ref{tt2}) is reversed with respect to the one of (\ref%
{tt}), which leads to instability (see the next section).

The result is confirmed at $p^{2}<0$, but with a caveat. Indeed, (\ref{ims})
is positive for $p^{2}<0$, but the cut diagram with propagators (\ref{tt2})
looks negative for both $p^{2}>0$ and $p^{2}<0$. The reason why there is no
contradition is that the cut diagrams give integrals that are divergent for $%
p^{2}<0$, so the final result may have an unexpected sign. If we use the
cutoff $\Lambda $ on the $k_{s}$ integral, it is easy to check (choosing
e.g. $p^{0}=0$) that the resul is (\ref{ims}) plus the linear divergence met
in formula (\ref{linear}). If we use the dimensional technique, the linear
divergences are set automatically to zero and we just find (\ref{ims}).

Diverging imaginary parts are quite unusual. However, they should be
expected when both prescriptions $\pm i\epsilon $ are present in the same
diagram, since in ref. \cite{ugo} it was found that both the locality and
hermiticity of counterterms are violated in such cases.

In the end, we do have a negative $\mathrm{Im}(-i\Sigma ^{\prime })$ for $%
p^{2}>0$, which means that the theory can be at most pseudounitary, but not
unitary. Since the right-hand side of the identity $-iT+iT^{\dag }=T^{\dag
}T $ cannot be positive definite, it should at best be replaced by $T^{\dag
}HT$, where $H$ is a matrix that has both positive and negative eigenvalues.
However, pseudounitarity is not a viable concept for a physically meaningful
theory.

In total, the imaginary part of $-2i\Sigma _{\text{FW}}$ follows from
formula (\ref{sfwres}) and is equal to%
\begin{equation}
2\mathrm{Im}(-i\Sigma _{\text{FW}})=\frac{\sqrt{u_{+}u_{-}}}{16\pi p^{2}}%
\left[ \theta (-u_{+})-\theta (u_{-})\right] .  \label{imsa}
\end{equation}%
It can be obtained as the sum of the cut diagrams of fig. \ref{cut} with one
cut propagator given by the top line of fig. \ref{cutprop} and the other cut
propagator given by the bottom line of that figure. Explicitly, one cut
diagram has propagators%
\begin{equation}
(2\pi )\theta (p^{0}-k^{0})\delta ((p-k)^{2}-m_{1}^{2}),\qquad \text{sgn}%
(k^{0})\pi \delta (k^{2}-m_{2}^{2}),  \label{tt3}
\end{equation}%
and the other one is obtained by flipping the signs of the arguments of the $%
\theta $ and sign functions.

If the FW\ particles were purely virtual, their cut propagator and $\mathrm{%
Im}(-i\Sigma _{\text{FW}})$ would have to vanish, to be consistent with the
optical theorem. Indeed, we know that the cut diagrams are related to the
cross section for the production of the particles circulating in the loop.
Purely virtual particles can never be produced, by definition. Since $%
\mathrm{Im}(-i\Sigma _{\text{FW}})$ is not zero, we must conclude that a FW\
particle is not purely virtual. Moreover, $\mathrm{Im}(-i\Sigma _{\text{FW}%
}) $ is nonvanishing even for $p^{2}<(m_{1}-m_{2})^{2}$. This result is
difficult to interpret, since it amounts to the production of particles with
negative energies and leads to instability (see the next section).

Note that formula (\ref{sfwfw}) gives the extremely simple result%
\begin{equation}
\Sigma _{\text{FW-FW}}(p)=-\mathrm{Im}(-i\Sigma _{\text{FW}})=-\frac{\sqrt{%
u_{+}u_{-}}}{32\pi p^{2}}\left[ \theta (-u_{+})-\theta (u_{-})\right] ,
\label{noana}
\end{equation}%
which means that the amplitude $-i\Sigma _{\text{FW-FW}}$ is purely
imaginary. Indeed, by definition $\Sigma _{\text{FW-FW}}$ is real.
Consistently with what we have found above, $2\mathrm{Im}(-i\Sigma _{\text{%
FW-FW}})$ is equal to the sum of the two cut diagrams of fig. \ref{cut} with
the cut propagators given by the third line of fig. \ref{cutprop}.

From the result (\ref{noana}) we see that the FW propagators break
analyticity in an unusual way. In section \ref{fakeons} we will see that,
instead, fakeons upgrade the common notion of analyticity to the more
general notion of regionwise analyticity.

In ref. \cite{bollini} the imaginary part of $\Sigma _{\text{FW}}(p)$ is
claimed to be just one half the imaginary part of the usual bubble diagram $%
\Sigma (p)$, since the authors miss the contribution from $\Sigma ^{\prime
}(p)$, which is the piece proportional to $\theta (u_{-})$ in formula (\ref%
{imsa}).

\section{Stability}

\setcounter{equation}{0}\label{stability}

In this section we show that a theory propagating FW particles violates
stability. Specifically, the denominators of (\ref{sp}) and (\ref{spr})
vanish for%
\begin{equation}
|p^{0}|=\omega _{1}-\omega _{2}.  \label{kine}
\end{equation}%
These singularities tell us where $\Sigma ^{\prime }(p)$ may be nonanalytic,
which in the case at hand means that it has branch cuts. The discontinuity
of the amplitude around the cuts leads to the nontrivial imaginary part $%
\mathrm{Im}(-i\Sigma ^{\prime })$. As we know, the optical theorem relates
the discontinuity to a physical process that turns the circulating virtual
particles into real particles.

The crucial aspect of formula (\ref{kine}) is that the external energy is
equated to a difference of frequencies, rather than a sum. Thus, if we
assume $p^{0}>0$, for definiteness, the physical process associated with $%
\mathrm{Im}(-i\Sigma ^{\prime })$ is an external particle of momentum $p$
decaying into a particle of energy $\omega _{1}$ and a particle of energy $%
p^{0}-\omega _{1}=-\omega _{2}$. This means instability, because the
energies of the final particles are unbounded.

Explicitly, for large momentum $k_{s}$, we have%
\begin{equation}
p^{0}=\omega _{1}-\omega _{2}\sim -p_{s}\cos \theta +\frac{1}{2k_{s}}\left[
m_{1}^{2}-m_{2}^{2}+p_{s}^{2}\sin ^{2}\theta \right] +\mathcal{O}\left( 
\frac{1}{k_{s}^{2}}\right) .  \label{conda}
\end{equation}%
Let us assume $m_{1}>m_{2}$ and take, for example, $\theta =\pi /2$. Then (%
\ref{conda}) has solutions for arbitrarily large $k_{s}$ and arbitrarily
small incoming energies $p^{0}$. If the incoming particle is massless, or
has a very small mass, it can decay into particles with arbitrarily large
positive and negative energies. Moreover, as we know from the divergent
behaviors of formulas (\ref{ims}), (\ref{imsa}) and (\ref{noana}) at $%
p^{2}=0 $, the probability of such an occurrence is huge and negative. This
is an unacceptable dynamics and another reason why the FW propagator is not
good for Feynman diagrams.

\section{Fakeons as purely virtual quanta}

\setcounter{equation}{0}\label{fakeons}

In this section we explain how the fakeons avoid the problems of the
Feynman-Wheeler particles and provide the correct concept of purely virtual
quanta.

The calculations of diagrams with fakeons require crucial modifications with
respect to the calculations of diagrams with Feynman propagators. For this
reason, it is convenient to briefly revisit the standard bubble diagram $%
\Sigma (p)$. Using Feynman parameters, the loop integral 
\begin{equation*}
\Sigma (p)=\int \frac{\mathrm{d}^{D}k}{(2\pi )^{D}}\hspace{0.01in}\frac{1}{%
(p-k)^{2}-m_{1}^{2}+i\epsilon }\,\frac{1}{k^{2}-m_{2}^{2}+i\epsilon }
\end{equation*}%
gives%
\begin{equation}
\Sigma (p)=V(p^{2}+i\epsilon ,m_{1}^{2},m_{2}^{2}),  \label{sigmap}
\end{equation}%
where 
\begin{equation}
V(a,b,c)=\frac{i\Gamma \left( \frac{4-D}{2}\right) }{(4\pi )^{D/2}}%
\int_{0}^{1}\mathrm{d}x\left[ -ax(1-x)+bx+c(1-x)\right] ^{(D-4)/2}.
\label{vabc}
\end{equation}%
Defining $u_{\pm }$ as in (\ref{upm}) and expanding around $D=4$, we obtain%
\begin{equation}
i(4\pi )^{2}V(p^{2},m_{1}^{2},m_{2}^{2})=-\ln \frac{4\Lambda ^{2}}{m_{1}m_{2}%
}+\frac{m_{1}^{2}-m_{2}^{2}}{p^{2}}\ln \frac{m_{1}}{m_{2}}-\frac{\sqrt{%
u_{+}u_{-}}}{p^{2}}\ln \frac{m_{1}^{2}+m_{2}^{2}-p^{2}+\sqrt{u_{+}u_{-}}}{%
2m_{1}m_{2}},  \label{sbubb}
\end{equation}%
where $\ln \Lambda ^{2}$ is defined again as in (\ref{cuto}).

Making (\ref{sbubb}) more explicit, we find \cite{hollik}%
\begin{eqnarray}
i(4\pi )^{2} &&V(p^{2}+i\epsilon ,m_{1}^{2},m_{2}^{2})=-\ln \frac{4\Lambda
^{2}}{m_{1}m_{2}}+\frac{m_{1}^{2}-m_{2}^{2}}{p^{2}}\ln \frac{m_{1}}{m_{2}} 
\notag \\
&&-\frac{\sqrt{u_{+}u_{-}}}{p^{2}}\theta (u_{-})\left( \ln \frac{\sqrt{u_{+}}%
+\sqrt{u_{-}}}{\sqrt{u_{+}}-\sqrt{u_{-}}}\right) +\frac{2\sqrt{-u_{+}u_{-}}}{%
p^{2}}\theta (-u_{-})\theta (u_{+})\arctan \sqrt{\frac{-u_{-}}{u_{+}}} 
\notag \\
&&+\frac{\sqrt{u_{+}u_{-}}}{p^{2}}\theta (-u_{+})\left( \ln \frac{\sqrt{%
-u_{-}}+\sqrt{-u_{+}}}{\sqrt{-u_{-}}-\sqrt{-u_{+}}}-i\pi \right) .
\label{hollik}
\end{eqnarray}

To explain how to proceed when the bubble diagram involves circulating
fakeons, we start from the case where one internal leg is a fakeon and the
other internal leg is a Feynman propagator. The loop integral reads%
\begin{equation*}
\Sigma _{\text{f}}(p)\equiv \int \frac{\mathrm{d}^{D}k}{(2\pi )^{D}}\hspace{%
0.01in}\frac{1}{(p-k)^{2}-m_{1}^{2}+i\epsilon }\,\left. \frac{1}{%
k^{2}-m_{2}^{2}}\right\vert _{\text{f}},
\end{equation*}%
where the fakeon prescription is denoted by means of the subscript f.

The simplest way to formulate the fakeon prescription is to make the Wick
rotation from the Euclidean version of the diagram and complete it
nonanalytically by means of an operation called average continuation \cite%
{LWformulation,fakeons}, which we describe below.

Specifically, we start from the Euclidean version of the diagram $\Sigma (p)$
and initiate the Wick rotation, i.e. move analytically from the Euclidean
region, where the energies are purely imaginary, to the Minkowskian region.
So doing, the thresholds we find on the real axis coincide with the ones of $%
\Sigma (p)$. In particular, the true threshold is still $%
p^{2}=(m_{1}+m_{2})^{2}$, while $p^{2}=(m_{1}-m_{2})^{2}$ is a
pseudothreshold. Below the threshold $p^{2}=(m_{1}+m_{2})^{2}$ we find no
obstacle and conclude the Wick rotation as usual. This means that the loop
integrals $\Sigma (p)$ and $\Sigma _{\text{f}}(p)$ coincide and are analytic
there:%
\begin{equation}
\Sigma _{\text{f}}(p)=\Sigma (p)\qquad \text{for }p^{2}<(m_{1}+m_{2})^{2}.
\label{below}
\end{equation}%
Note that the fact that the pseudothreshold does not behave differently from
usual is crucial for stability, since kinematic relations like (\ref{kine})
do not play any role.

Above the threshold $p^{2}=(m_{1}+m_{2})^{2}$ the difference between $\Sigma
_{\text{f}}(p)$ and $\Sigma (p)$ becomes apparent. In the case of $\Sigma
(p) $, the threshold is crossed analytically by means of the Feynman
prescription $p^{2}\rightarrow p^{2}+i\epsilon $. This means that the Wick
rotation is completed analytically. In the case of $\Sigma _{\text{f}}(p)$
the threshold is crossed by means of the fakeon prescription, which amounts
to taking the arithmetic average of the two analytic continuations around
the threshold, which correspond to the prescriptions $p^{2}\rightarrow
p^{2}\pm i\epsilon $. This operation, called average continuation \cite%
{LWformulation,fakeons}, is unambiguous, but not analytic. Hence, we speak
about nonanalytic Wick rotation \cite{LWformulation}. Despite being
nonanalytic, it returns an analytic function above the threshold. What is
not analytic is just the relation between the two analytic functions that
encode the amplitude below and above the threshold.

We know that $\Sigma (p)$ is given by formula (\ref{hollik}). When we
replace a particle circulating in the loop with a fakeon, the only
difference is that the $i\pi $ of the last line disappears. Indeed, the $%
i\pi $ of (\ref{hollik}) is due to having crossed the threshold by means of
the analytic continuation $p^{2}\rightarrow p^{2}+i\epsilon $. The other
analytic continuation $p^{2}\rightarrow p^{2}-i\epsilon $ gives $-i\pi $, so
the arithmetic average of the two gives zero. At the end, equation (\ref%
{below}) extends to%
\begin{eqnarray}
\Sigma _{\text{f}}(p) &=&\frac{1}{2}\Big(V(p^{2}+i\epsilon
,m_{1}^{2},m_{2}^{2})+V(p^{2}-i\epsilon ,m_{1}^{2},m_{2}^{2})\Big)  \notag \\
&=&\frac{1}{2}\Big(\Sigma (p)-\Sigma ^{\ast }(p)\Big)=\Sigma (p)+\frac{\sqrt{%
u_{+}u_{-}}}{16\pi p^{2}}\theta (-u_{+}).  \label{above}
\end{eqnarray}%
In particular,%
\begin{equation}
2\mathrm{Im}(-i\Sigma _{\text{f}})=0.  \label{unit}
\end{equation}%
This result is consistent with the optical theorem. We know that the
discontinuity of the amplitude above the threshold, which is encoded in its
imaginary part, is associated with a physical process where the particles
circulating in the loop become real. However, a purely virtual quantum
cannot become real, by definition. This means that when fakeons are involved
such a process has zero chances to occur, i.e. the imaginary part vanishes
above the threshold. Then the potential physical process becomes a fake
process.

If we interpret (\ref{unit}) in terms of the cutting equations, so that $2%
\mathrm{Im}(-i\Sigma _{\text{f}})$ equals the sum of the cut diagrams of
fig. \ref{cut}, we can say that the cut propagator of the fakeon is
identically zero, as shown in fig. \ref{cutprop}.

Analyticity holds above $p^{2}=(m_{1}+m_{2})^{2}$ in both cases $\Sigma $
and $\Sigma _{\text{f}}$, but in different senses. In the case of $\Sigma _{%
\text{f}}$, the space of complexified external momenta $p$ is divided into
two disjoint regions of analyticity. One region is located below the
threshold and the other one is located above the threshold. The former is
the Euclidean region, the latter is a fakeon region.

Specifically, if we take formula (\ref{hollik}) and drop the $i\pi $
appearing in the last line, we have an analytic function for $%
p^{2}<(m_{1}+m_{2})^{2}$ and another analytic function for $%
p^{2}>(m_{1}+m_{2})^{2}$. The two are not analytically related to each
other. However, the former unambiguously determines the latter by means of
the average continuation.

Note that, in general, there is no way to determine the function below the
threshold from the function above the threshold \cite{fakeons} (although in
the particular case at hand it seems that we may achieve this goal). For
example, in three spacetime dimensions we often meet the square-root
function $\sqrt{z}$ ($z$ being $-p^{2}-i\epsilon $ or a more complicated
function of $p^{2}$). Its average continuation on the negative real axis is
zero and obviously we cannot recover $\sqrt{z}$ from the zero function.

The new analyticity property of the amplitudes involving fakeons is called
\textquotedblleft regionwise analyticity\textquotedblright .

Note that the fakeon prescription is free of nonlocal divergences. Indeed,
the divergences coincide with those of $\Sigma (p)$, which are local.

If both particles circulating in the loop are fakeons, we need to compute
the loop integral%
\begin{equation*}
\Sigma _{\text{ff}}(p)\equiv \int \frac{\mathrm{d}^{D}k}{(2\pi )^{D}}\hspace{%
0.01in}\left. \frac{1}{(p-k)^{2}-m_{1}^{2}}\right\vert _{\text{f}}\,\left. 
\frac{1}{k^{2}-m_{2}^{2}}\right\vert _{\text{f}}.
\end{equation*}%
We can proceed as above, starting from the Euclidean framework and ending
the Wick rotation nonanalytically by means of the average continuation above
the threshold $p^{2}=(m_{1}+m_{2})^{2}$. The result is%
\begin{equation}
\Sigma _{\text{ff}}(p)=\Sigma _{\text{f}}(p)=\Sigma (p)+\frac{\sqrt{%
u_{+}u_{-}}}{16\pi p^{2}}\theta (-u_{+})  \label{sff}
\end{equation}%
and again $2\mathrm{Im}(-i\Sigma _{\text{ff}})=0$.

We can summarize the crucial differences among the calculations of $\Sigma
(p)$, $\Sigma _{\text{f}}(p)$, and $\Sigma _{\text{FW}}(p)$ as follows. $%
\Sigma (p)$ can be computed directly in Minkowski spacetime or by means of
the analytic Wick rotation from the Euclidean version of the loop integral; $%
\Sigma _{\text{f}}(p)$ is computed from the Euclidean version, but the Wick
rotation is completed nonanalytically by means of the average continuation; $%
\Sigma _{\text{FW}}(p)$ is computed directly in Minkowski spacetime, since
the principal-value prescription is inherently Minkowskian. The integral on
the loop energy picks the same residues in the cases of $\Sigma (p)$ and $%
\Sigma _{\text{f}}(p)$, but different residues in the case of $\Sigma _{%
\text{FW}}(p)$.

In the end, the results obtained with the fakeon propagator are quite
different from the ones due to the FW propagator. With fakeons: ($i$) the
optical theorem, hence unitarity, holds; ($ii$) no instability is generated,
since the thresholds are the same as usual, which correspond to kinematics
like%
\begin{equation*}
|p^{0}|=\omega _{1}+\omega _{2},
\end{equation*}%
instead of (\ref{kine}); ($iii$) finally, there is no problem with the
locality of counterterms, since the divergent part can be computed in the
Euclidean region, where, by formula (\ref{below}), $\Sigma _{\text{f}}$ and $%
\Sigma $ coincide. Once the amplitude is renormalized there, its average
continuation is renormalized everywhere, like its analytic continuation.

\section{Lower dimensions}

\setcounter{equation}{0}

\label{three}

Now we repeat the analysis of FW particles in three and two spacetime
dimensions.

In three dimensions we assume $p^{2}>0$ and make the calculation by adapting
formula (\ref{sp}). We obtain%
\begin{equation*}
\Sigma ^{\prime }(p)=\frac{i}{8\pi \sqrt{p^{2}}}\ln \frac{\sqrt{%
p^{2}-i\epsilon }+m_{1}-m_{2}}{\sqrt{p^{2}+i\epsilon }-m_{1}+m_{2}}.
\end{equation*}%
The (negative) imaginary part%
\begin{equation*}
2\mathrm{Im}(-i\Sigma ^{\prime })=-\frac{\theta (u_{-})}{4\sqrt{p^{2}}}
\end{equation*}%
can be verified by evaluating the cut diagrams of fig. \ref{cut} with the
rules explained in section \ref{unitarity}.

In two dimensions it is more convenient to use the dimensional
regularization and expand the result of (\ref{uabc}) around $D=2$. We find%
\begin{equation*}
U(p^{2},m_{1}^{2},m_{2}^{2})=\frac{i}{4\pi \sqrt{u_{+}u_{-}}}\ln \frac{1+z}{%
1-z}.
\end{equation*}%
Thus, using (\ref{sempla}) the discontinuity of the amplitude is%
\begin{equation*}
2\mathrm{Im}(-i\Sigma ^{\prime })=-\frac{\theta (u_{-})}{\sqrt{u_{+}u_{-}}},
\end{equation*}%
which is also negative. Again, the result can be verified by computing the
cut diagrams of fig. \ref{cut} with the rules of section \ref{unitarity}.

We see that the problems found in four dimensions are essentially confirmed
in lower dimensions.

\section{Fakeons, God's energy and the infinite desert}

\setcounter{equation}{0}

\label{remarks}

We point out that the fakeon prescription virtualizes a particle (or a
ghost) completely and eradicates it from the theory. In particular, a theory
with fakeons is as fundamental as the standard model.

If a theory has ghosts, it is not acceptable as a fundamental theory, even
if the ghosts have a finite lifetime and decay. Indeed, an unstable particle
or ghost is not really out of the physical spectrum, in the same way as the
muon is not out of the physical spectrum of the standard model.
\textquotedblleft Living with ghosts\textquotedblright\ \cite{hawking} is
not a viable option, even if the ghosts may be unobservable in common
settings.

The only wayout is to quantize the would-be ghosts in a radically different
way, that is to say as fakeons. Then they are really out of the physical
spectrum, at all energies.

The most important application of the fakeon prescription is that it allows
us to make sense of quantum gravity as a perturbative quantum field theory 
\cite{LWgrav,UVQG}. The theory is described by the action 
\begin{equation}
S_{\text{QG}}(g,\Phi )=-\frac{1}{2\kappa ^{2}}\int \mathrm{d}^{4}x\sqrt{-g}%
\left[ 2\Lambda _{C}+\zeta R+\alpha \left( R_{\mu \nu }R^{\mu \nu }-\frac{1}{%
3}R^{2}\right) -\frac{\xi }{6}R^{2}\right] +S_{\mathfrak{m}}(g,\Phi ),
\label{SQG}
\end{equation}%
where $\alpha $, $\xi $, $\zeta $ and $\kappa $ are positive constants, $M_{%
\text{Pl}}=1/\sqrt{G}=\sqrt{8\pi \zeta }/\kappa $ is the Planck mass, $\Phi $
are the matter fields and $S_{\mathfrak{m}}$ is the action of the matter
sector. Besides the graviton, the theory propagates two massive fields: a
scalar $\phi $ and a spin-2 field $\chi _{\mu \nu }$. The residue of the $%
\chi _{\mu \nu }$ free propagator has the wrong sign, so $\chi _{\mu \nu }$
must be quantized as a fakeon, because the Feynman prescription would turn
it into a ghost. The residue at the $\phi $ pole has the correct sign, so $%
\phi $ can be quantized either as a fakeon or a true particle. The masses $%
m_{\phi }=\sqrt{\zeta /\xi }$ and $m_{\chi }=\sqrt{\zeta /\alpha }$
(neglecting small corrections due to the cosmological constant $\Lambda _{C}$%
) are free parameters and should be determined experimentally. Their values
could be smaller, or even much smaller, than the Planck mass, e.g. $m_{\phi
}\sim m_{\chi }\sim $10$^{12}$GeV. The fine structure constants that govern
the perturbative expansion of (\ref{SQG}) are the ratios $\alpha _{\chi
}=m_{\chi }^{2}/M_{\text{Pl}}^{2}$ and $\alpha _{\phi }=m_{\phi }^{2}/M_{%
\text{Pl}}^{2}$ and could be as small as $10^{-14}$ \cite{absograv}.

These facts imply that the pure gravitational sector of the theory is
perturbative up to an unbelievably high energy. Indeed, since the action (%
\ref{SQG}) is renormalizable, the perturbative expansion is not governed by
the ratio $E/M_{\text{Pl}}$, where $E$ is the center-of-mass energy, but by
the running couplings $\alpha _{\chi }$ and $\alpha _{\phi }$. The running,
in turn, is governed by logarithmic corrections. It takes a long way to turn
the products 
\begin{equation*}
\alpha _{\chi }\ln \frac{E^{2}}{\mu ^{2}},\qquad \alpha _{\phi }\ln \frac{%
E^{2}}{\mu ^{2}},
\end{equation*}%
into quantities of order 1, whatever reference energy $\mu $ we take.
Precisely, $\alpha _{\chi }\sim \alpha _{\phi }\sim 10^{14}$ and $\mu \sim
M_{\text{Pl}}$ give the unbelievably high energy%
\begin{equation*}
E\sim 10^{10^{13}}M_{\text{Pl}},
\end{equation*}%
which we think deserves to be called \textquotedblleft God's
energy\textquotedblright\ for this reason. If we could multiply the maximum
energy we can reach in our laboratories by a factor 10 every year, we would
reach the Planck scale in less than twenty years and God's energy in one
thousand times the age of the universe.

Note that small couplings $\alpha _{\chi }$ and $\alpha _{\phi }$ do not
mean that the theory is practically free. For example, $\phi $ and $\chi
_{\mu \nu }$ have widths that are proportional to their masses times $\alpha
_{\phi }$ and $\alpha _{\chi }$, respectively \cite{absograv}. Since the
masses are large, we obtain nonnegligible widths even if $\alpha _{\phi }$
and $\alpha _{\chi }$ are small. Actually, the width of $\chi _{\mu \nu }$
is comparable to the widths of the $Z$ and $W$ bosons and the one of the
Higgs boson $H$.

Of course, when we say that the theory is perturbative up to God's energy,
we refer to the elementary processes (graviton-graviton scattering,
graviton-matter scattering, etc.). Nonperturbative problems are present at
all energies, when they involve large numbers of particles and gravitons at
the same time, as in the classical limit, black holes, etc.

The quantum gravity theory based on the fakeon idea predicts new physics
below the Planck scale (precisely, at energies equal to the masses of $\phi $
and $\chi $) and then at the Planck scale itself. At the same time, it opens
up an extremely alarming scenario, which is the threat of an
\textquotedblleft infinite desert\textquotedblright\ from the Planck scale
to God's energy: no new physics, nothing interesting, forever! Although we
can always assume that extra, very heavy particles and/or fakeons exist in
such a huge range of energies without affecting any fundamental principles,
this kind of variety might be rather unexciting.

Long ago Weinberg proposed asymptotic safety as a way to overcome the
nonrenormalizability of Einstein gravity \cite{wein}. The idea relies on the
assumption that the ultraviolet limit is an interacting conformal fixed
point with a finite-dimensional critical surface. In Weinberg's approach one
has to advocate nonperturbative or semi-nonperturbative methods, which make
it difficult to discuss the issue of unitarity. Moreover, assumptions on the
ultraviolet limit are hard to accept, given that it is experimentally out of
reach, by definition.

A related issue is ultraviolet completeness, which is defined in different
ways and sometimes linked to asymptotic safety or asymptotic freedom,
although such notions do not appear to be necessary requirements for
completeness. We emphasize that the theory based on the fakeon idea (which
is not asymptotically free \cite{UVQG,betas}) is safe enough and ultraviolet
complete enough, due to its huge perturbative regime. Finally, its
nonperturbative sector is candidate to explain even what lies beyond God's
energy. In this sense, it can be considered ultraviolet complete.

\section{Conclusions}

\setcounter{equation}{0}

\label{conclusions}

The possibility that purely virtual entities might exist in nature is
interesting in itself and has attracted the attention of several scientists
in the past, both at the classical and quantum levels. However, the correct
candidate was not identified right away. At the classical level, Dirac
virtualized runaway solutions by renouncing causality at small distances,
while Feynman and Wheeler considered T-symmetric wave emissions in classical
electrodynamics. At the quantum level, Bollini and Rocca picked up on the
suggestion of Feynman and Wheeler and studied the Cauchy principal value as
a propagator in Feynman diagrams.

For a variety of reasons, having both prescriptions $\pm i\epsilon $ in the
same diagram is extremely dangerous, because it leads to violations of the
locality of counterterms, unitarity and stability. The right purely virtual
quanta turn out to be the fakeons, which can be used to virtualize both
ghosts and normal particles. We have compared the Feynman, fakeon and
Feynman-Wheeler prescriptions in the bubble diagram and studied the optical
theorem in four, three and two dimensions. Only the fakeon has an
identically vanishing cut propagator, which means that it remains virtual
once the quantum corrections are turned on.

Fakeons allow us to make sense of quantum gravity at the fundamental level,
and place that theory on an equal footing with the standard model. The
theory of quantum gravity that emerges from the fakeon idea is perturbative
up to an unbelievably high energy, which we named God's energy. This opens
up a frightening scenario: the possibility of an infinite desert between the
Planck scale and God's energy, with no new physics in sight, basically
forever.

\vskip 12truept \noindent {\large \textbf{Acknowledgments}}

\vskip 2truept

I am grateful to U. Aglietti for helpful discussions. I also thank the
participants of the conferences \href{http://www.noncommutativegeometry.nl/qgqg2019/}%
{\textquotedblleft Quantum gravity and quantum geometry\textquotedblright },
Radboud University, Nijmegen, Oct.-Nov., 2019, and \textquotedblleft \href{https://www.perimeterinstitute.ca/conferences/cosmological-frontiers-fundamental-physics-2019}%
{Cosmological frontiers in fundamental} \href{https://www.perimeterinstitute.ca/conferences/cosmological-frontiers-fundamental-physics-2019}%
{physics 2019}\textquotedblright , Sept. 2019, Perimeter Institute, for
valuable exchanges.

\end{document}